\documentclass[12pt]{article}
\usepackage{amsfonts,amssymb,amsmath}
\usepackage[dvips]{epsfig}
\begin{document}
\title{\bf Generalized Coherent States for the Spherical
Harmonics $Y_{m}^{m}(\theta,\phi)$}
\author{H. Fakhri\thanks{Email: hfakhri@tabrizu.ac.ir}
\\
{\small {\em Department of Theoretical Physics and
Astrophysics, Physics Faculty,}}\\
{\small {\em University of Tabriz, P O Box 51666-16471, Tabriz, Iran}}\\ \\
B. Mojaveri\thanks{Email: bmojaveri@raziu.ac.ir}\\
{\small {\em Department of Physics, Azarbaijan Shahid Madani University,}}\\
{\small {\em Tabriz 53741-161, Iran}}}
\maketitle
\begin{abstract}
\noindent
The associated Legendre functions $P_{l}^{(m)}(x)$ for a given $l-m$, may be taken into account as the
increasing infinite sequences with respect to both indices $l$ and $m$.
This allows us to construct the exponential generating functions
for them in two different methods by using Rodrigues formula. As an application then
we present a scheme to construct generalized coherent states corresponding to the spherical
harmonics $Y_{m}^{m}(\theta,\phi)$.
\\ {\bf MSC Nos:} 33C45; 05A15
\end{abstract}
\section{Introduction}
Coherent states were first discovered by Schr\"odinger in the
context of quantum mechanics in order to minimize the uncertainty relation between
momentum and position  coordinates \cite{Schrodinger}, and were later generalized
successfully to the Lie group approaches, by
Glauber, Klauder, Sudarshan, Barut, Girardello and Perelomov
\cite{Glauber1,Glauber2,Klauder1,Klauder2,Sudarshan,Barut,Perelomov1}.
Also, for the models with one degree of freedom either discrete or continuous
spectra -with no remark on the existence of a Lie algebra
symmetry- Gazeau and Klauder proposed new coherent states which
are parameterized by two real parameters
\cite{Gazeau,Antoine,Fakhri1}.
The quantum coherency of states has
nowadays pervaded many branches of physics such as quantum optics,
quantum electrodynamics, solid-state physics, nuclear and atomic
physics, from both theoretical and experimental viewpoints \cite{Klauder3,Perelomov2,Klauder4,Ali}.
Moreover, some generalized approaches
in connection with coherent states corresponding to
shape invariant models have been proposed \cite{Fukui,Fakhri2}.

Any infinite superposition from the pure states of a quantum
mechanical system cannot form coherent states because they not
only must be converged to the finite-valued functions but also
must accept a positive definite measure to satisfy the resolution
of the identity condition on the entire complex plane or on a
unit disc. When both of them are satisfied, then they, as
coherent superpositions, minimize uncertainty relation for some
values of the complex variable. Using the Barut-Girardello
eigenvalue equation for the laddering operators of $su(1,1)$ Lie
algebra, the coherent states corresponding to the spherical
harmonics $Y_{m}^{m}(\theta,\phi)$ have been calculated
\cite{Fakhri3}. Also, in Ref. \cite{Fakhri4}, we have shown that
the use of the spatial parity symmetry for
$Y_{m}^{m}(\theta,\phi)$ as angular wave functions of the
one-partite systems can lead to entangled
$su(1,1)$-Barut-Girardello coherent states for a bipartite quantum
system. In this paper for the first time, we introduce exponential
and non-exponential generating functions corresponding to the
associated Legendre functions. Then, we construct the
generalized coherent states of  $Y_{m}^{m}(\theta,\phi)$ as an
application to the new generating functions.

\section{New generating functions for the associated Legendre functions}
A large number of physical and chemical contexts involve application  of the
associated Legendre functions $P_{l}^{(m)}(x)$.
They are given by the Rodrigues formula \cite{Bateman,Wang,Vilenkin} ($-1<x<+1$)
\begin{eqnarray}
&&\hspace{-18mm}P_{l}^{(m)}(x)=a_{l}^{(m)}
(1- x^2)^{-\frac{m}{2}}
\left(\frac{d}{dx}\right)^{l-m} \,\left(1-x^2\right)^{l},
\end{eqnarray}
in which, $l$ is a non-negative integer and $m$, an integer number, is bounded by $l$: $-l \leq m \leq l$.
The normalization coefficients are
\begin{eqnarray}
a_{l}^{(m)}=\left\{\begin{array}{l}\frac{(-1)^{l}}{2^{l} \,l!}\,
\frac{(l+m)!}{(l-m)!}\hspace{20mm} 0\leq m\leq l,
\\ \\ \frac{(-1)^{l-m}}{2^{l} \,l!} \hspace{25mm}-l\leq m\leq 0.
\end{array}\right.
\end{eqnarray}
Also, the functions with positive and negative values of $m$
are proportional with each other
\begin{eqnarray}
P_{l}^{(-m)}(x)=(-1)^m\frac{(l-m)!}{(l+m)!}P_{l}^{(m)}(x),
\end{eqnarray}
whilst the associated Legendre functions with the same $l-m$ ($l+m$)
but with different values of $l$ and $m$ are independent of each other.

Thus, we can introduce two different types of infinite sequences of
the associated Legendre functions depending
on whether $l-m$ is  even or odd.
These types of sequences are increasing
with respect to both indices $l$ and $m$ of the functions.
Due to the fact that whether $l-m$ is even or odd, i.e. $l-m=2k$ or
$l-m=2k+1$ with $k$ as a non-negative integer, the lowest functions are
$P_{k}^{(-k)}(x)$ and $P_{k+1}^{(-k)}(x)$, respectively.
It is obvious that the terminology of lowest functions is devoted to the associated
Legendre functions $P_{l}^{(m)}(x)$ with the lowest value for both indices $l$ and $m$.
For a given value of $k$, the generating functions corresponding to the
sequences are calculated as
\begin{eqnarray}
&&\hspace{-8mm}G_{k}^{\mbox{{\tiny even}}}(x,t)=\sum_{m=-k}^{\infty}\frac{t^{k+m}}{(k+m)!}
\frac{P_{2k+m}^{(m)}(x)}{a_{2k+m}^{(m)}}\nonumber\\
&&\hspace{10.5mm}=\sum_{m=0}^{\infty}\frac{t^{m}}{m!}(1-x^2)^{\frac{k-m}{2}}
\left(\frac{d}{dx}\right)^{2k}\left(1-x^2\right)^{k+m}
\nonumber\\
&&\hspace{10.5mm}=\sum_{m=0}^{\infty}\frac{t^{m}}{m!}(1-x^2)^{\frac{k-m}{2}}\frac{(2k)!}{2\pi i}
\oint_{C} dz \frac{(1-z^{2})^{k+m}}{(z-x)^{2k+1}}
\nonumber\\
&&\hspace{10.5mm}=(1-x^2)^{\frac{k}{2}}\frac{(2k)!}{2\pi i}
\oint_{C} dz \frac{(1-z^{2})^{k}exp\left(\frac{t(1-z^2)}{\sqrt{1-x^2}}\right)}{(z-x)^{2k+1}}
\nonumber\\
&&\hspace{10.5mm}=(1-x^2)^{\frac{k}{2}}\left[\left(\frac{d}{dz}\right)^{2k}
\left((1-z^{2})^{k}exp\left(\frac{t(1-z^2)}{\sqrt{1-x^2}}\right)\right)\right]_{z=x},\\
&&\hspace{-8mm}G_{k}^{\mbox{{\tiny odd}}}(x,t)=\sum_{m=-k}^{\infty}\frac{t^{k+m}}{(k+m)!}
\frac{P_{2k+m+1}^{(m)}(x)}{a_{2k+m+1}^{(m)}}\nonumber\\
&&\hspace{10.5mm}=\sum_{m=0}^{\infty}\frac{t^{m}}{m!}(1-x^2)^{\frac{k-m}{2}}
\left(\frac{d}{dx}\right)^{2k+1}\left(1-x^2\right)^{k+m+1}
\nonumber\\
&&\hspace{10.5mm}=\sum_{m=0}^{\infty}\frac{t^{m}}{m!}(1-x^2)^{\frac{k-m}{2}}\frac{(2k+1)!}{2\pi i}
\oint_{C} dz \frac{(1-z^{2})^{k+m+1}}{(z-x)^{2k+2}}
\nonumber\\
&&\hspace{10.5mm}=(1-x^2)^{\frac{k}{2}}\frac{(2k+1)!}{2\pi i}
\oint_{C} dz \frac{(1-z^{2})^{k+1}exp\left(\frac{t(1-z^2)}{\sqrt{1-x^2}}\right)}{(z-x)^{2k+2}}
\nonumber\\
&&\hspace{10.5mm}=(1-x^2)^{\frac{k}{2}}\left[\left(\frac{d}{dz}\right)^{2k+1}
\left((1-z^{2})^{k+1}exp\left(\frac{t(1-z^2)}{\sqrt{1-x^2}}\right)\right)\right]_{z=x},
\end{eqnarray}
for $\left|t\right|<\infty$.
$C$ is a closed contour in positive direction on the complex plane $z$.
In order to satisfy Eqs. (4) and (5), it is sufficient that the arbitrary contour $C$ is chosen so that
the point $z=x$ lies inside of that.
The accordance of the above generating functions with
the theorem 1 of Ref. \cite{Srivastava} can be considered as confirmation for it.
Therefore, depending on whether $l-m$ is  even or odd integer, we have obtained
two different types of generating functions as  multipliers of $exp(t\sqrt{1-x^2})$
for the associated Legendre functions. Note that
Eq. (3) does not allow us to get new generating functions by means of the sequences with given values of $l+m$.

Furthermore, it should be emphasized that by using the closed
contour applied to Jacobi polynomials, new generating functions
corresponding to infinite sequences of the functions with the same
(non-negative) $m$ are derived as
\begin{eqnarray}
&&\hspace{-8mm}G_{m}(x,t)=\sum_{l=m}^{\infty}\frac{t^{l-m}}{(l-m)!}
\frac{P_{l}^{(m)}(x)}{2^{l}a_{l}^{(m)}}=
\frac{\left(\frac{-1-xt+\sqrt{t^2+2xt+1}}{t^2\sqrt{1-x^2}}\right)^m}{\sqrt{t^2+2xt+1}},
\hspace{17mm} \left|t\right|<1.
\end{eqnarray}
In the special case $m=0$, (6) is converted to the known generating function of the Legendre polynomials.

\section{Generalized coherent states for the spherical
harmonics $Y_{m}^{m}(\theta,\phi)$}
This section covers the method of making generalized coherent states for the spherical
harmonics $Y_{m}^{m}(\theta,\phi)$ by using the new generating functions of the associated
Legendre functions. The spherical harmonics are described in terms of the polar (or co-latitude)
angle $0\leq \theta \leq \pi$
and the azimuthal (or longitude) angle $0\leq \phi<2\pi$:
\begin{eqnarray}
\left|l,m\right\rangle:=Y_{l}^{m}(\theta,\phi)=\sqrt{\frac{2l+1}{4\pi}\frac{(l-m)!}{(l+m)!}}\,e^{im\phi}P_{l}^{(m)}(\cos\theta).
\end{eqnarray}
They form an orthonormal set with respect to the following inner product
\begin{eqnarray}
\left\langle l,m|l^{\prime},m^{\prime}\right\rangle
:=\int_{\theta=0}^{\pi}\int_{\phi=0}^{2\pi}{Y_{l}^{m}}^{*}(\theta,\phi)Y_{l^{\prime}}^{m^{\prime}}(\theta,\phi)
d\Omega(\theta,\phi)=\delta_{l\,l^{\prime}}\delta_{m\,m^{\prime}}.
\end{eqnarray}
For a given $l-m=2k$ with the lower bound $m\geq -k$, the
generating function obtained in (4) can be used to construct
coherent states as follows: Let us define the infinite-dimensional
Hilbert space ${\cal
H}_k:=\mbox{span}\left\{\left|2k+m,m\right\rangle\right\}_{\hspace{1mm}
m\geq -k}$ equipped with the identity operator
$\sum_{m=-k}^{\infty}\left|2k+m,m\right\rangle\left\langle
2k+m,m\right|=I_{_{{\cal H}_k}}$. As an infinite superposition of
the spherical harmonics, generalized coherent states together with
their explicit compact forms can be calculated by (4) as
\begin{eqnarray}
&&\hspace{-10mm}\left|z\right\rangle_k:=N_k(|z|)
\sum_{m=-k}^{\infty}\frac{(2k+m)!}{(k+m)!}\frac{z^m}{\sqrt{(4k+2m+1)(2k+2m)!}}\left|2k+m,m\right\rangle \nonumber\\
&&\hspace{-1mm}=N_k(|z|)
\frac{\left(\frac{-\sin\theta}{2e^{i\phi}z}\right)^k}{\sqrt{4\pi(2k)!}}
\left[\frac{d^{2k}}{du^{2k}}
\left(\left(1-u^2\right)^{k}e^{-\frac{z\left(1-u^2\right)}{2\sin\theta}e^{i\phi}}\right)\right]_{u=\cos\theta},
\end{eqnarray}
in which $N_k(|z|)$ are the the normalization coefficients.
Here, $z$ is an arbitrary complex variable with the polar form $z=re^{i\varphi}$ so that
$0\leq r<\infty$ and $0\leq \varphi<2\pi$. The
main ingredient of this work is the convergence of the infinite
expansions $\left|z\right\rangle_k$ as coherent states to explicit
compact forms of the known functions.

In what follows it is assumed that $k=0$. If the norm of the
coherent states $\left|z\right\rangle$ is supposed to be
normalized to unity with respect to the inner product (8), i.e.,
$\left\langle  z|z\right\rangle=1$, then we find the explicit form
$N(|z|)=\sqrt{\frac{|z|}{\sinh|z|}}$ for the real normalization
coefficient. Also, we should introduce the appropriate measure
$d\mu(|z|)=rK(r)drd\varphi$ so that the resolution of the identity
is realized for the coherent states
\begin{eqnarray}
&&\hspace{-10mm}\left|z\right\rangle=\sqrt{\frac{|z|}{4\pi\sinh|z|}}e^{-\frac{z}{2}e^{i\phi}\sin\theta}
\end{eqnarray}
in the Hilbert space ${\cal H}_0$,
\begin{eqnarray}
&&\hspace{-7mm}I_{_{{\cal H}_0}}=
\int_{\mathbb{C}}|z\rangle\langle z|\,d\mu(|z|)
=2\pi\sum_{m=0}^{\infty}\frac{\left|m,m\right\rangle\left\langle  m,m\right|}{(2m+1)!}\int_{0}^{\infty}r^{2m+1}N^2(r)K(r)dr.
\end{eqnarray}
Using the completeness relation
$\sum_{m=0}^{\infty}\left|m,m\right\rangle\left\langle m,m\right|=I_{_{{\cal H}_0}}$
it is found that relation (11) is satisfied for the  positive definite measure
$K(r)=\frac{\sinh r}{2\pi r}e^{-r}$.

\section{Conclusions}
The parameter $m$ allows us to calculate two different and new types of
exponential generating functions (4) and (5) for the associated Legendre functions.
Also, generating function corresponding to the Legendre polynomials
can be obtained as a special case of the non-exponential generating functions (6)
of the associated Legendre functions.
The generating function corresponding to infinite sequence $\{P_m^{(m)}(x)\}_{m=0}^{\infty}$
of the associated Legendre functions is used as an application example to construct the generalized
coherent superposition $\left|z\right\rangle$ of the spherical harmonics $Y_{m}^{m}(\theta,\phi)$.
Its explicit compact form and also, to realize the
resolution of the identity, its corresponding positive definite measure on the
complex plane have been calculated.

\end{document}